\def\beg{\begin{equation}}
\def\eeq{\end{equation}}
\documentstyle[12pt]{article}
\begin{document}
\begin{center}
{\Large{\bf Disappearance of fractional statistics in Schrieffer-Wilczek theory.}}
\vskip0.35cm
{\bf Keshav N. Shrivastava}
\vskip0.25cm
{\it School of Physics, University of Hyderabad,\\
Hyderabad  500046, India}
\end{center}

The paper of Arovas, Schrieffer and Wilczek is corrected. It is 
found that the Berry's phase is vanishingly small. Accordingly, 
the statistical vector potential also becomes negligibly small.
The need for intermediate statistics is then obviated.
\vfill
Corresponding author: keshav@mailaps.org\\
Fax: +91-40-3010145.Phone: 3010811.
\newpage
\baselineskip22pt
\noindent {\bf 1.~ Introduction}

     As the magnetic field is varied, the Hall conductance, $\sigma_{xy}$= $I_x/E_y$=$\nu e^2/h$ shows plateaus at $\nu$=$n/m$
where $n$ and $m$ are integers, with $m$ being odd. Laughlin$^1$ 
has shown that at the plateaus, the quasiparticle charge in the ``incompressible fluid" is $\pm e^*$=$e/m$. Arovas et al$^2$ make 
an effort to determine the statistics of the quasiparticles of 
fractional charge by determining the Berry's phase in the ``incompressible fluid".

In this letter, we report that the calculation of the phase for incompressible fluid has not been carried out correctly by Arovas 
et al$^2$ and if performed correctly will not give the same result. 
In particular, the Berry's phase found by Arovas et al
is large but the correct value is vanishingly small. Similarly, 
the correction to the vector potential found by Arovas et al$^2$ 
is large but when calculated properly, becomes negligibly small.
Therefore, the need for intermediate statistics is completely 
obviated.

\noindent{\bf 2.~~Theory}

      Arovas et al report that the quasiparticle charge is $\pm e^*$=$\pm e/m$. Actually, Laughlin found that the charge density is $\sigma_m$=1/[m(2$\pi a_o^2$)] where $a_o^2$=$\hbar c/eB$. On the basis of this result Laughlin introduced incompressibility so that $a_o$ can be left out as a fixed number and the charge is determined to be $e/m$. This result is in error because there are at least three possibilities. (i) Keep the charge at $e/m$ and a$_o^2$ as a constant, (ii) fix the charge at $e$ and change the area from $a_o^2$ to $ma_o^2$ and (iii) the charge is $e$, the area is a$_o^2$ and $m$ is kept as a number. Laughlin has chosen only the first of these possibilities. Hence the question is that what scientific basis has 
been used to pick up (i) and leave out (ii) and (iii)? The question 
is about the selection of (i) by ignoring (ii) and (iii). There is 
obviously no reason to prefer (i) and there is no method which 
dictates that only (i) is the correct answer. In fact, a linear combination of all the three should be chosen if the correct answer 
is to be obtained. In a product of $e/ma_o^2$  it is not correct to say that the charge $e$ has been changed to $e/m$ because it is also possible that $e$ is unchanged and only $a_o^2$ is changed to $ma_o^2$. Laughlin's choice is therefore arbitrary. We wish to emphasize that there are other solutions which were left out by Laughlin. Arovas et al take $\pm e^*$=$\pm e/m$ but obviously the 
correct result is not the charge $e^*$ but the charge density, $e/2m\pi a_o^2$ which is the charge density per unit area. It is possible to consider the one dimensional result, $e/ma_o$, which is 
the charge per unit length or the three dimensional result, i.e., 
the charge per unit volume, e/[m(4$\pi a_o^3$/3)]. We shall consider only the charge per unit area. Once the result for the area is known, the length or the volume results can be trivially derived.

     Arovas et al calculate the change of phase, $\gamma$ of a state
having a quasihole localized at $z_o$. As $z_o$ adiabatically moves around a circle of radius $R$ enclosing flux $\phi$, the change in 
phase is,
\beg
{e^*\over\hbar c}\oint \vec A.d\vec l= 2\pi({e^*\over e})({\phi\over \phi_o}).
\eeq
This is the phase gained by the quasiparticle of charge $e^*$ in 
going around the loop. Actually, the Laughlin calculation is 
concerned with the charge per unit area and not only the 
quasiparticle charge, so that $e^*$ should be replaced by 
$e^*/ma_o^2$. {\bf Arovas et al have left out the factor $ma_o^2$ 
by using the argument of incompressibility but $a_o$ involves the 
magnetic field which causes field dependence}. Since, $e^*$
is the charge in a magnetic field, it uses $a_o^2$=$\hbar c/eB$ 
to determine the relevant area. Therefore, the charge $e$ in the 
absence of a magnetic field is linked to the Bohr radius, $a_B$.
The corrected form of the phase must have $ma_o^2$. Since the phase 
must be dimensionless, the corrected form of the phase should be,
\beg
\gamma =2\pi ({e^*\over e})({a_B^2\over ma_o^2})({\phi\over \phi_o}).
\eeq
This means that the phase factor is linearly dependent on the 
magnetic field. At the fields of a few Tesla, $a_o\sim 10^{-4} 
\,\,cm$ while $a_B\sim 10^{-8}\,\, cm$ so that $(a_B/a_o)^2 \simeq
 10^{-8}$. Therefore, the phase factor $\gamma$ is a negligibly small number  or it can be said that no phase is gained. On the other hand, Arovas et al claim that there is a large gain. Since they have left 
out the variables, it is clear that they have over estimated the 
phase. The usual phase is $\omega t$ which is modified to $\omega t + \gamma $. If the area $ma_o^2$ is taken into account, $\gamma$ tends 
to zero. If $\omega$ is the frequency and $\pi/a_o$ is the wave vector, then the velocity is $\omega a_o/\pi$ and $\gamma$ is not quite independent of the velocity. If $z_o$ is integrated in a clockwise 
sense around a circle of radius $R$, the values of $|z|<R$
contribute $2\pi i$ to the integral. Therefore,
\beg
\gamma_o = 2\pi \nu \phi/\phi_o.
\eeq
Comparing this with (1),
\beg
\nu e = e^*.
\eeq
Correcting the effective charge for the charge density,
\beg
{e\over ma_o^2} = {\nu e\over a_B^2}
\eeq
we find,
\beg
\nu = {a_B^2 \over ma_o^2}.
\eeq
If $a_B = a_o$, then $\nu = 1/m$. For $\phi = \phi_o$,
\beg
\gamma_o = 2 \pi \nu
\eeq
or the corrected value is
\beg
\gamma_o^{corr} = 2\pi \nu (a_B^2/a_o^2).
\eeq
Since, $a_B << a_o, \gamma_o^{corr}$ tends to zero. Hence, the 
phase pickup tends to zero. If there is a finite phase, then the 
vector potential $\vec A_o$ should be corrected by adding to it a ``statistical" vector potential $A_{\phi}$ such that, 
\beg
{e^*\over \hbar c}\oint {\vec A_{\phi}}.d{\vec l} = \Delta\gamma=2\pi\nu.
\eeq

When above corrections are considered, this $\vec A_{\phi}$ tends 
to zero. As long as (7) is used, the phase for interchanging quasiparticles is $\Delta\gamma/2=\pi$. That will change the sign 
for Fermi statistics. If $\nu$ is a fraction, then the quasiparticles cease to be interchanged like fermions so they obey the ``fractional statistics".    We have lost contact with the anticommutators but 
have not gained the commutators. Therefore,the particles are neither fermions nor bosons for certain fractional values of the phase. 
When $A_{\phi}$ tends to zero, we need not determine the statistics.
When $\nu \to$ 0,  Arovas et al suggest the factor of $\phi_o$
for bosons and $\phi_o(1-1/\nu)$ for fermions, but as $\nu \to 0$
there is a divergence in $\phi_o$. Obviously, Wilczek and Zee$^3$
prescription requires to be corrected. Some authors put the 
correction in the vector potential without correcting the scalar potential. Such a procedure can not give the correct ``Chern-Simons" contribution. In the case of fermions, the interchange of 
quasiparticles produces a minus sign or a factor of -1, whereas in the case of bosons the required factor is +1. Therefore, the antisymmetry
 produces a factor of -1. In the case of product of fermion operators
the factor produced will be $\pm1$ but never and never 3, then why Laughlin requires it to be 3? So, it will never be 3. Hence, Laughlin also requires to be modified. In the case of Peierls distortion, Su 
and Schrieffer$^4$ suggest many manipulations of charge. In fact they devide the charge as $e/2$ and $e/2$ but no care is taken to devide 
the mass of the electron. When there is a Peierls distortion at twice the lattice
constant, the wave vector becomes $\pi/2a$ and it is suggested that charge be counted as $e/2$ and $e/2$. It is more likely that charge 
will devide as 0 and $e$ and not as $e/2$ and $e/2$. Similarly, if 
there are trimers, Su and Schrieffer$^4$ suggest $e/3$ but here also
no suggestion is given to devide the mass of the electron, so the 
charge may devide as 0, 0, and $e$ and not as $e/3$, $e/3$ and $e/3$.
It is obvious that Su and Schrieffer's suggestions are at best qualitative and not found in real TTF-TCNQ or CH$_x$ type compounds.
It is true that Jackiw and Rebbi$^5$ do find ${1\over 2}$ of a fermion
number by Dirac equation but not by Peierls distortion.

\noindent{\bf 3.~~Conclusions}

     The best one can say is that Su and Schrieffer's suggestions are qualitative and not found in nature. Similarly, the ideas of fractional statistics are not set on good foundation. Even if the 
system is ``incompressible" the results of Laughlin are not 
sufficiently rigorous and at best are arbitrary.

     There has been a lot of drum beating in favour of composite fermions (CF), which have the flux quanta attached to the electron to modify the magnetic field but the flux attached electrons will be heavier than found in the experimental data by several orders of magnitude$^6$. Similarly, the CF are much bigger objects than electrons to have the same density. Hence the CF model is also not correct.
                                                                                  
The correct theory of the quantum Hall effect is given in ref.7

\noindent{\bf6.~~References}
\begin{enumerate}
\item R. B. Laughlin, Phys. Rev. Lett. {\bf50}, 1395(1983).
\item D. Arovas, J. R. Schrieffer and F. Wilczek, Phys. Rev. Lett. {\bf53}, 722 (1984).
\item F. Wilczek and A. Zee, NSF-ITP-84-25 (1984).
\item W. P. Su and J. R. Schrieffer, Phys. Rev. Lett. {\bf46}, 738 (1981).
\item R. Jackiw and C. Rebbi, Phys. Rev. D {\bf13}, 3398 (1976).
\item K. N. Shrivastava, cond-mat/0210320.
\item K. N. Shrivastava, Introduction to quantum Hall effect,\\ 
      Nova Science Pub. Inc., N. Y. (2002).
\end{enumerate}
\vskip0.1cm
Note: Ref.7 is available from:\\
 Nova Science Publishers, Inc.,\\
400 Oser Avenue, Suite 1600,\\
 Hauppauge, N. Y.. 11788-3619,\\
Tel.(631)-231-7269, Fax: (631)-231-8175,\\
 ISBN 1-59033-419-1 US$\$69$.\\
E-mail: novascience@Earthlink.net

\vskip0.5cm

\end{document}